\documentclass[conference]{IEEEtran}
\IEEEoverridecommandlockouts
\usepackage{cite}
\usepackage{amsmath,amssymb,amsfonts}
\usepackage{algorithmic}
\usepackage{graphicx}
\usepackage{textcomp}
\usepackage{xcolor}
\usepackage{multirow}
\usepackage{url}
\usepackage{listings}
\def\BibTeX{{\rm B\kern-.05em{\sc i\kern-.025em b}\kern-.08em
    T\kern-.1667em\lower.7ex\hbox{E}\kern-.125emX}}

\begin{document}

\title{Comparing Two Counting Methods for Estimating the Probabilities of Strings}
% {\footnotesize }}

\author{\IEEEauthorblockN{\small{Ayaka Takamoto\quad Mitsuo Yoshida\quad Kyoji Umemura}}
\IEEEauthorblockA{\textit{\small{Department of Computer Science and Engineering}}\\
\textit{\small{Toyohashi University of Technology}}\\
\small{Toyohashi, Aichi, Japan} \\
\small{takamoto.ayaka.nx@tut.jp, yoshida@cs.tut.ac.jp, umemura@tut.jp}}
}

% \IEEEpubid{978-1-6654-1743-3/21/$31.00~\copyright~2021 IEEE$}
\IEEEpubid{\makebox[\columnwidth]{978--1--6654--1743--3/21/\$31.00~\copyright~2021 IEEE \hfill} \hspace{\columnsep}\makebox[\columnwidth]{ }}
\maketitle

\begin{abstract}
There are two methods for counting the number of occurrences of a string in another large string.
One is to count the number of places where the string is found.
The other is to determine how many pieces of string can be extracted without overlapping.
The difference between the two becomes apparent when the string is part of a periodic pattern.
This research reports that the difference is significant in estimating the occurrence probability of a pattern.

In this study, the strings used in the experiments are approximated from time-series data. The task involves classifying strings by estimating the probability or computing the information quantity.
First, the frequencies of all substrings of a string are computed. 
Each counting method may sometimes produce different frequencies for an identical string.
Second, the probability of the most probable segmentation is selected.
The probability of the string is the product of all probabilities of substrings in the selected segmentation.
The classification results demonstrate that the difference in counting methods is statistically significant, and that the method without overlapping is better.

\end{abstract}
\vspace{\baselineskip}
\renewcommand\IEEEkeywordsname{Keywords}
\begin{IEEEkeywords}
\textit{Information quantity, Probability estimation, Frequency, String statistics, String overlaps}
\end{IEEEkeywords}

\section{Introduction}
The classification of strings is a classical and important task\cite{viterbi1967error}.
A reasonable approach involves estimating the probability of a given string by assuming that the string originates from each class, and then, selecting the class that gives the highest probability.
In this approach, the probability estimation method determines the classification effectiveness.

The problem of zero-frequency occurs for long strings.
For example, for a given page of a book, it is unlikely that there exists an identical sequence of words on a page in any other book.
In natural language, the set of words on a page can be used to find documents that are similar to that page; however, this is difficult for a general string.
Takamoto et al.~\cite{takamoto2017computing} proposed using the information quantity of the string to classify musical scores by composer.
Their approach determines the most probable segmentation of data, and subsequently, uses the product of the probability of all strings in the segment.
The most frequent substring then becomes the natural clue for classification.
Thus, this solves the problem of zero-frequency.
In this approach, it is important to choose an appropriate technique for estimating the probability.

When a string is part of a repetitive pattern, the correct number of string occurrences is not apparent.
For example, there are two counting methods for the occurrences of ``abca'' in ``abcabcabca.''
The first method is overlapping counting.
This method searches for ``abca'' and finds three instances of ``abca.''
Hence, the ``a'' in the middle belongs to two found strings and overlaps in the two strings.
The other method is non-overlapping counting. 
After the first ``abca'' is found in ``abcabcabca,'' it is removed, and the next ``abca'' is found from only ``bcabca.''
Using this method,  two occurrences of ``abca'' are found in ``abcabcabca.'' 

\begin{figure}[t]
  \centerline{\includegraphics[width=8cm]{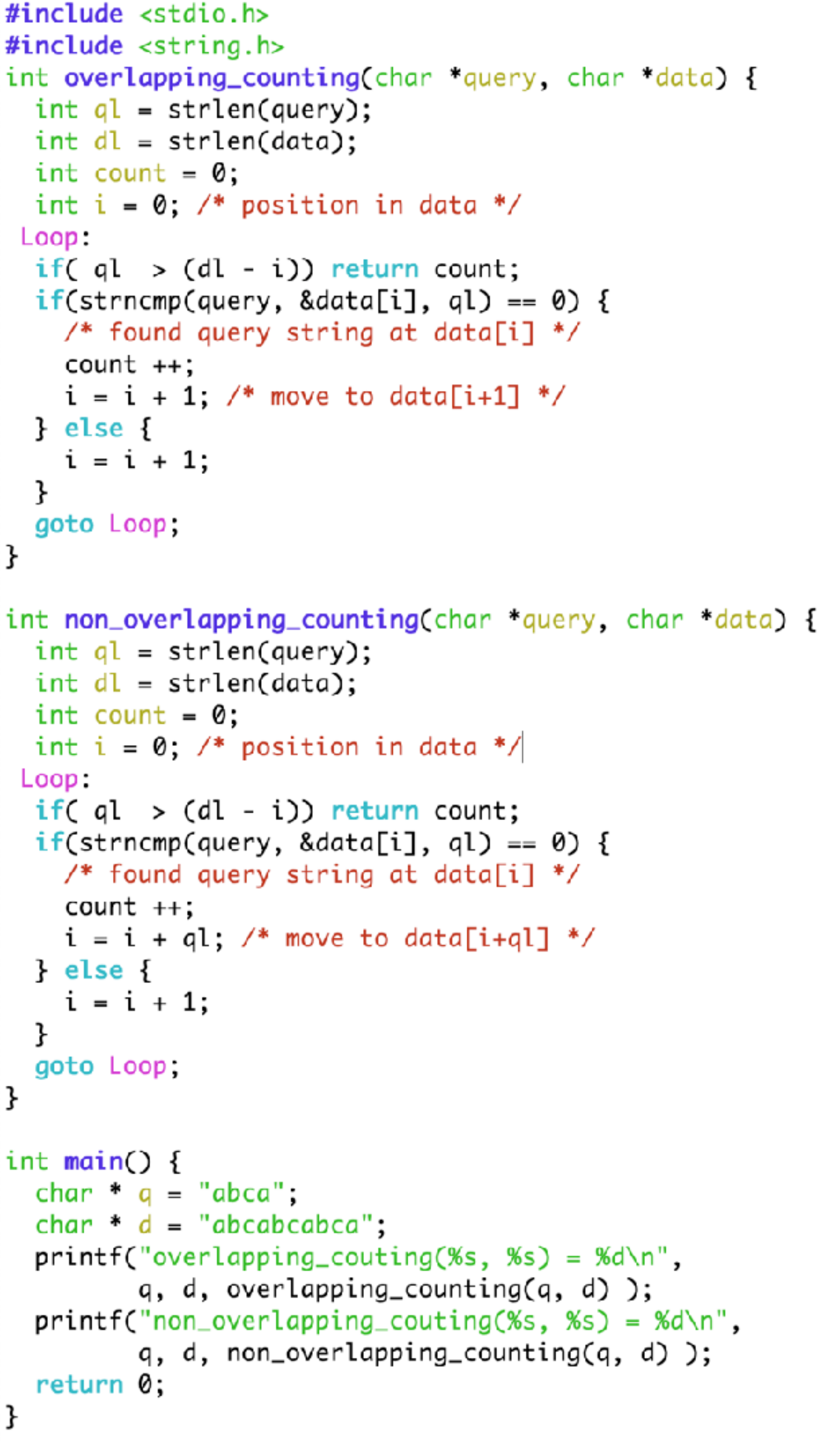}}
\caption{C code for two counting methods. Overlapping counting always moves to the next character for the next search, whereas non-overlapping counting skips the matched string for the next search.}
\label{fig_code}
\end{figure}

The two counting methods are defined by the C code shown in Fig.~\ref{fig_code}.
The function counts the query string in the data string.
If the data string is shorter than the query string, the counting result is zero for both methods.
If the first part of the data string matches the query string, the counting value is incremented.
The difference between the two counts is the start point for the next search.
Non-overlapping disregards the matched part, whereas overlapping starts from the next character.
If the first part of the data does not match the query string, both start searching from the next character.

\begin{figure}[t]
\centerline{\includegraphics[width=9.5cm]{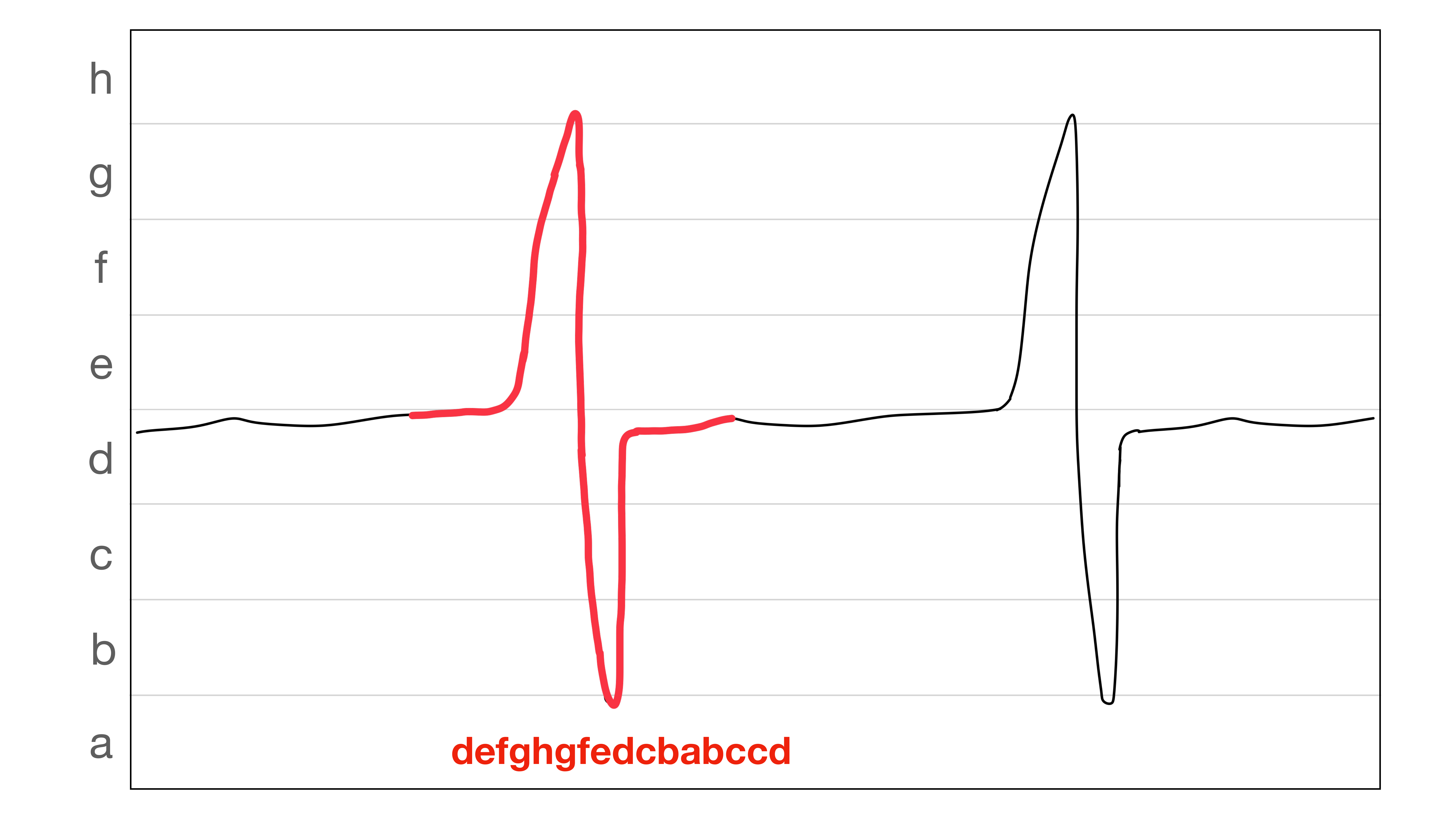}}
\caption{Query pattern comprising a pair of waves. The string below the curve corresponds to the string representation of a single wave. The corresponding character is assigned to a wave, according to the range shown on the vertical axis.
}
\label{fig1}
\end{figure}

\begin{figure}[t]
\centerline{\includegraphics[width=8.5cm]{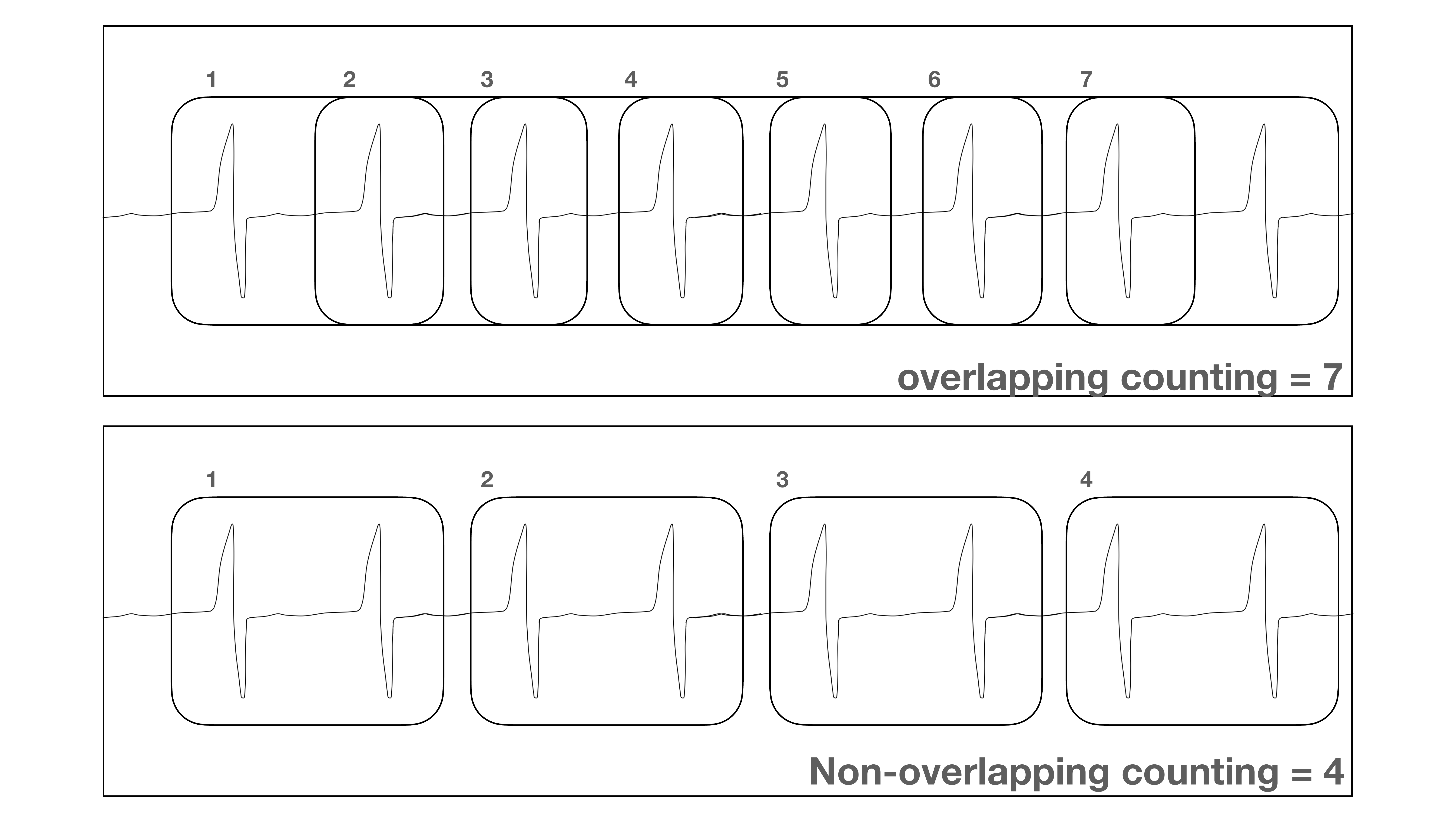}}
\caption{Two different answers to the question, “How many query patterns exist in the sequence of eight waves.'' The answer from overlapping counting is seven, whereas the answer from non-overlapping counting is four.}
\label{fig2}
\end{figure}

The string classification method can be applied to time-series data, including sound wave data, using string approximation.
Time-series data can be efficiently processed via string approximation.
However, the overlap in the first method may cause problems in the time-series analysis using string approximation.
Consider a query sound of 2ms, comprising two waves, as shown in Fig.~\ref{fig1}.
Furthermore, consider the following question: “How many query sounds exist in 8ms of sound comprising the same eight waves?”
The natural answer is four, as shown in the lower part of Fig.~\ref{fig2}.
Note also that there are seven timing positions where two waves exist in a row within 8ms, as shown in the upper part of Fig.~\ref{fig2}.
It is difficult to determine which of two or seven is more appropriate.

\IEEEpubidadjcol

In this research, the strings in the experiments are approximated from time-series data.
The task involves classifying strings by estimating the probability or computing the information quantity.
First, the frequencies of all substrings of a string were computed. Each counting method may produce different frequencies for an identical string.
Second, the probability of the most probable segmentation is selected.
The probability of the string is the product of all probabilities of substrings in the selected segmentation.
The classification result shows that the difference in counting methods is statistically significant, and that the method without overlapping is better.

\section{Time-series Data Classification Method}

\subsection{Symbolic Aggregate Approximation}
In the first step, a sequence of floating-point numbers is converted into a string.
Time-series data are usually expressed as a sequence of floating-point numbers. 
The symbolic aggregate approximation~(SAX) was proposed by Lin et al.~\cite{lin2003symbolic} for this purpose.
This research uses SAX as the first step. 
SAX translates the data into character strings, as an approximated form of data.
It first assigns some range of floating pointer numbers to one character.
This approximation is called piecewise aggregate approximation.
Suppose that there are time-series data of size $n$, that is, comprising $n$ floating-point numbers.
The data are first approximated (downsampled) into a $w$ floating-point number, where $w\le n$.
Then, the range is decided such that the number of resulting floating-point numbers is equal.
Using this range, the resulting floating pointer is assigned to a character. 
The original time-series data of size $n$ become a string of length $w$.

\subsection{Estimating probability of long string}
The estimating method used herein is the same as that used by Takamoto et al.~\cite{takamoto2017computing}.
For the given test string and class, the frequency in the class is computed for all substrings of the test string.
Then, the corresponding estimator of probability is computed for all substrings of the test string.
Using this probability, the more probable segmentation of the test string is decided.
The probability is the product of the probability of each string in the segmentation.
This is expressed by the following formula, wherein information quantity is used instead of the probability.

\begin{equation}
    I_s(T_1; T_2) = \min_{\pi_k \in \pi(T_1)}\left(-\sum_{t \in \pi_k}\log_2P(t; T_2)\right)
\end{equation}
where $T_1$ is the test string, and $T_2$ is a class expressed by a string.
$\pi(T_1)$ is the set of all partitions of string $T_1$.
The size of this set is $2^{length(T_1)-1}$. $P(t; T_2)$ is the estimated probability of $t$, provided that $t$ is drawn from $T_2$.
The actual program uses dynamic programming or Vitabi Search~\cite{viterbi1967error} to ensure that the computation time is reasonable.
Using this definition, certain frequent substrings are considered as units.

We are predominantly interested in estimating $p(t; T_2)$. The apparent method is to use the most likely estimator (MLE), expressed as follows:

\begin{equation}
    \hat{P}(t; T_2) = \frac{{\rm freq}(t; T_2)}{|T_2|}.
\end{equation}
where $freq(t; T_2)$ is the number of times that $t$ appears in string $T_2$.
If $t$ is a single character or the length is always one, there is no ambiguity. However, if $t$ is part of a repetitive pattern, there is a difference between the overlapping and non-overlapping counts.

The definition and procedure for the time-series data follow Takamoto et al.~\cite{takamoto2017computing}, where the information quantity for music scores was computed.
Moreover, although Takamoto et al.~\cite{takamoto2017computing}  uses overlapping counting, non-overlapping counting has not been used for string classification based on information quantity.
This is because the overlapping count is efficient, and the difference between the two counting methods is not well known.

\begin{figure}[t]
\centerline{\includegraphics[width=9cm]{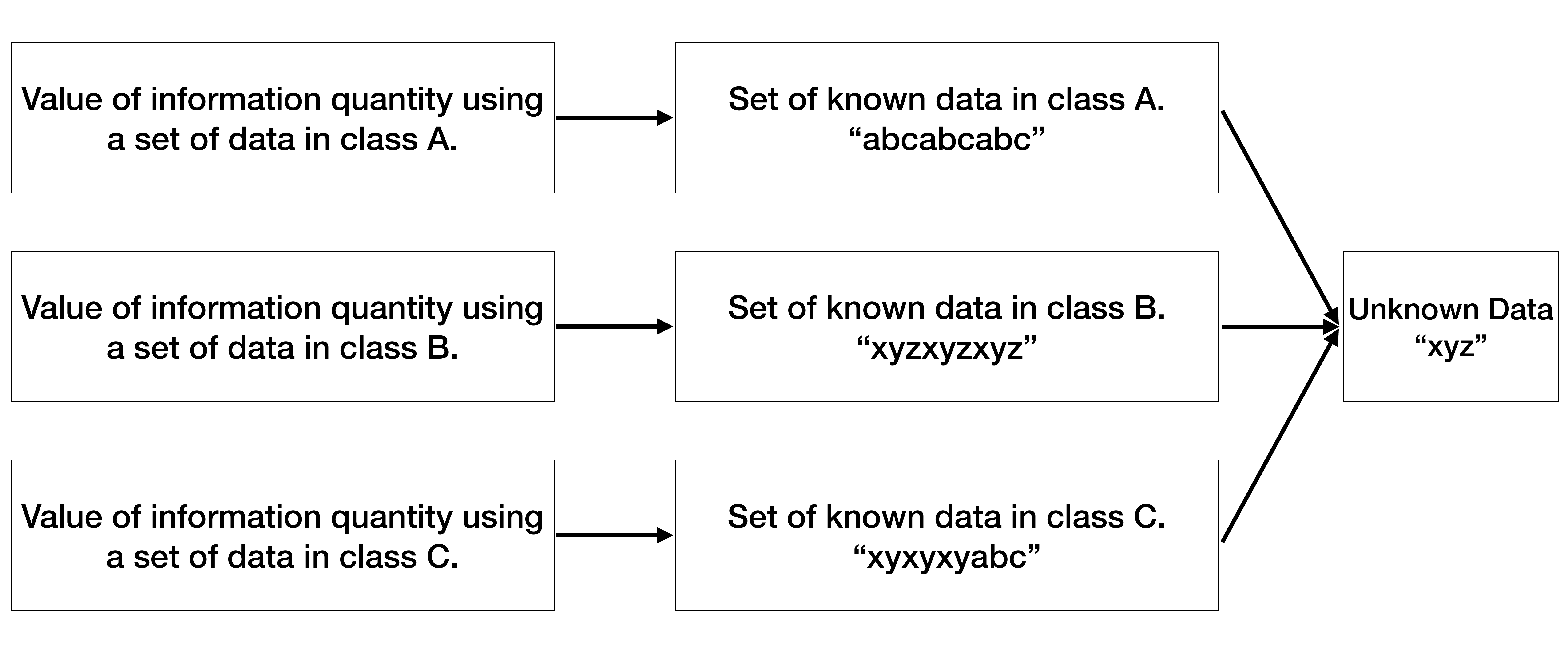}}
\caption{For the test string “xyz'' whose class is unknown, each class, A, B, and C, is considered. 
Because the class string “xyzxyzxyz'' produces the minimum information quantity (maximum probability), the estimated class will be B.}
\label{fig}
\end{figure}

\section{Experiment}
\subsection{Data}
The data in our experiment were downloaded from the website USR time series~\cite{Chen2015}.
Each set of time-series data comprises a label, which indicates the class for the data, and a sequence of numerical values. One dataset comprises training and test data.
Each time-series dataset contains the same sequence length, and is of a similar type (indicated by the data name) but different (indicated by class).
For example, the dataset ``Gun\_Point'' contains the moving angles of a finger.
It has two classes, i.e., one corresponding to gun operation, and the other to a finger pointing without a gun.
Table~\ref{tab:data} lists all names of the set of time-series data in our experiment.

\subsection{Preparation}
For each time-series dataset, the sequence of numerical values was converted into a string using SAX.
Sixteen alphabets, from ``a'' to ``p,'' are assigned to some range of floating-point numbers so that each alphabet is nearly equal in the dataset.
The length of the string is equal to the number of numerical values, that is, $w = n$.
Then, for each class in the dataset, a relatively long string (class string) is prepared by selecting and combining strings in the training dataset for the class.

\subsection{Classification}
First, either the overlapping or non-overlapping counting  method is selected to estimate the probability.
For each time-series dataset, a string (test string) is selected from every test set.
Then, the information quantity of the test string is computed using every class string.
The estimated class of the test string is that whose class string produces the minimum information quantity (or maximum probability).
%The example is shown in Fig.~

\subsection{Evaluation}
First, the number of correct estimations is used to evaluate each counting method.
This is presented in Table ~\ref{tab:data}. In Table ~\ref{tab:data}, the one that uses overlapping counting is labeled by a previous study, and the one that uses non-overlapping counting is labeled by the proposed method.
Both methods function similarly. The dataset name is underlined if non-overlapping counting performs better.
Notably, for the dataset with a large number of test sets, non-overlapping counting seems to achieve a better estimation.

Next, the two counting methods are compared using all 50 datasets.
Because the two methods function similarly, it is reasonable to compare Cases 2 and 3 using the McNamer test.
The results of all test strings in all datasets were assigned to one of four cases.
\begin{itemize}
    \item Case 1: Both overlapping counting and non-overlapping yields correct estimation.
    \item Case 2: Only overlapping counting yields correct estimation.
    \item Case 3: Only non-overlapping counting yields correct estimation.
    \item Case 4: Neither overlapping counting nor non-overlapping counting yield correct estimation.
\end{itemize}
The number of times each of the four cases occurs was counted.
The results are presented in Table~\ref{McNemer_test}.
There are 45660 test strings in the 50 dataset.
The numbers in Cases 1 and 4 are ignored because the number does not contain information on the difference between the two counting methods.
The difference between Cases 2 and 3 shows the difference in the effectiveness of the two counting methods. The null hypothesis is that both methods are equally effective.
Under this hypothesis, the number of Cases 2 and 3 is the result of equal probability.
This difference is statistically tested using the McNamer test.
Suppose Case 2 is $b$ times and Case 3 is $c$ times, then the critical ratio of the McNamer test is given as follows: 

\begin{equation}
    p=(\frac{1}{2^{b+c}}\sum_{k=0}^bC_k).
\end{equation}

In our study, $b = 623$, and $c = 1264$. The critical ratio was approximately $2.22 × 10^{-50}$.
This indicates that the difference was statistically significant. 
The number of test strings (45660) was sufficiently large to show statistical significance.
\begin{table}[t]

\begin{center}

\caption{Number of correct estimation by overlapping or non-overlapping counting}
\label{McNemer_test}
\scalebox{0.9}{
\begin{tabular}{c c | c c |c}
\hline
& & \multicolumn{2}{|c|}{Non-overlapping count} & \\
& & Correct result & Incorrect result & Total \\
\hline

\multirow{2}{*}{Overlapping count} & Correct result & 26049 & 623 & 26672 \\
 &Incorrect result & 1264 & 17724 & 18988 \\

\hline
& Total & 27313 & 18347 & 45660 \\
\hline
\end{tabular}
}
\end{center}
\end{table}
\section{Discussion}
\subsection{Relationship to compression-based dissimilarity measure}
Time-series datasets were prepared to show the performance of the compression-based dissimilarity measure (CDM), proposed by Keogh et al. \cite{keogh2004towards}.
The CDM for given string $x$ and $y$ is defined as follows.
\begin{equation}
    CDM(x, y) = \frac{C(xy)}{C(x) + C(y)}
\end{equation}
where $C(z)$ is the compressed size of string $z$, and $xy$ is the concatenated string of $x$ and $y$.
The assumption of this measure is that, if $x$ and $y$ are structurally similar, the compressed size of the concatenated string is smaller than the sum of the individually compressed sizes.
In the classification task, after computing the CDMs between the target and each training data, the estimated class would be that of the training data that gives the minimum CDM value. 

Theoretically, the information quantity of a string is the lower bound of its compressed size. This is expressed as follows:
 $I^*_s(t; t) < C(t)$,  where $I^*_s(t; t)$ is the true value of $I_s(t; t)$.

Although it is possible to use $I_s(t; t)$ instead of $C(t)$, an estimation error may become large when length $t$ is insufficient.
Therefore, $I_s(t; class\_string)$ will be estimated more accurately than $I_s(t; one\_training\_data\_string)$, as $class\_string$ is a long string (it is the concatenation of all training data strings).

\subsection{When should non-overlapping counting be used?}
The difference in the effectiveness of the methods is subtle. 
Although the experiment in this study shows a clear difference between the two methods, both work reasonably well.
This explains why non-overlapping counting has not been studied well to date.
Nevertheless, non-overlapping counting seems to be a natural counting method when probability is discussed.
For the question ``How many character pairs exist in the string of length 10?'' The natural answer was 5.
Then, for the question ``How many {``aa''} exist in {``aaaaaaaaaa''}?'' The natural answer was 5, rather than 9.

\subsection{Is there any efficient algorithm for non-overlapping count?}
Notably, after preparation, overlapping counting is an efficient use of data structures, such as the suffix array ~\cite{manber1993suffix} or suffix tree~\cite{weiner1973linear}.
Using the relationship reported by Umemura et al.~\cite{umemura2018non}, it is possible to use the suffix array to speed up non-overlapping counting; however, the overlapping count is less efficient.
Notably, there is a paper regarding the formation of a table of non-overlapping counts using an extended suffix tree~\cite{brodalsolving}.
We attempted to implement this algorithm; however, efficient implementation was found to be difficult. Currently, efficiency remains an issue for non-overlapping counting.

\section{Conclusion}
This study compares overlapping and non-overlapping counting aimed at time-series analysis.
The most important issue is estimating the probability or computing the quantity of information.
First, the frequencies of all substrings of the string were computed. Each counting method sometimes produces a different frequency for an identical string.
Second, the probability of the most probable segmentation is selected.
The probability of the string is the product of all probabilities of substring in the selected segmentation.
The classification results show that both counting methods function similarly, but with a slight difference.
The McNamer test shows that the difference in counting methods is statistically significant, and that the method without overlapping is better.

\begin{table*}[tb]
\begin{center}
\caption{Dataset information and number of correct answers}
\scalebox{1.25}{

\begin{tabular}{|l|r|r|r|r|r|}
\hline
Data Name&Number of&Size of&Data&Overlapping&Non-overlapping\\
&Classes&Testing Set&Length&Counting&Counting\\
\hline
\underline{50words}&50&455&134&147&154\\
Beef&5&30&234&17&17\\
BeetleFly&2&20&255&14&14\\
BirdChicken&2&20&255&13&12\\
CBF&3&900&63&581&537\\
Car&4&60&287&43&43\\
\underline{ChlorineConcentration}&3&3840&82&2269&2288\\
\underline{CinC\_ECG\_torso}&4&1380&818&611&784\\
Computers&2&250&359&150&148\\
\underline{Cricket\_X}&12&390&149&120&122\\
\underline{Cricket\_Y}&12&390&149&116&118\\
\underline{Cricket\_Z}&12&390&149&122&125\\
\underline{DiatomSizeReduction}&4&306&171&269&270\\
DistalPhalanxOutlineAgeGroup&3&400&39&332&332\\
\underline{ECG5000}&5&4500&69&3981&3989\\
ECGFiveDays&2&861&67&617&609\\
FISH&7&175&230&124&124\\
\underline{FaceAll}&14&1690&64&752&754\\
FaceFour&4&88&174&40&40\\
\underline{FacesUCR}&14&2050&64&946&959\\
\underline{Gun\_Point}&2&150&74&136&139\\
\underline{Haptics}&5&308&545&102&103\\
Herring&2&64&255&34&34\\
\underline{InlineSkate}&7&550&940&117&142\\
\underline{InsectWingbeatSound}&11&1980&127&390&450\\
\underline{LargeKitchenAppliances}&3&375&359&260&284\\
MALLAT&8&2345&511&2014&2000\\
Meat&3&60&223&57&57\\
\underline{MedicalImages}&10&760&48&382&383\\
\underline{MiddlePhalanxOutlineAgeGroup}&3&400&39&288&290\\
\underline{MoteStrain}&2&1252&41&1064&1067\\
\underline{NonInvasiveFatalECG\_Thorax1}&42&1965&374&1065&1179\\
\underline{NonInvasiveFatalECG\_Thorax2}&42&1965&374&1326&1417\\
OliveOil&4&30&284&26&26\\
OSULeaf&6&242&212&144&143\\
\underline{Phoneme}&39&1896&511&284&309\\
RefrigerationDevices&3&375&359&190&190\\
\underline{ScreenType}&3&375&359&178&188\\
SonyAIBORobotSurface&2&601&34&386&381\\
Strawberry&2&613&116&575&575\\
SwedishLeaf&15&625&63&408&406\\
\underline{Symbols}&6&995&198&528&608\\
\underline{Trace}&4&100&136&95&96\\
\underline{TwoLeadECG}&2&1139&40&896&906\\
\underline{Two\_Patterns}&4&4000&63&1820&1849\\
\underline{WordsSynonyms}&25&638&134&188&196\\
\underline{WormsTwoClass}&5&181&449&111&114\\
Worms&2&181&449&87&84\\
\underline{synthetic\_control}&6&300&29&174&178\\
yoga&2&3000&212&2083&2080\\
\hline
\end{tabular}
}
\label{tab:data}
\end{center}
\end{table*}

\vspace{12pt}
\bibliographystyle{IEEEtran}
\bibliography{reference}

\begin{thebibliography}{1}
\providecommand{\url}[1]{#1}
\csname url@samestyle\endcsname
\providecommand{\newblock}{\relax}
\providecommand{\bibinfo}[2]{#2}
\providecommand{\BIBentrySTDinterwordspacing}{\spaceskip=0pt\relax}
\providecommand{\BIBentryALTinterwordstretchfactor}{4}
\providecommand{\BIBentryALTinterwordspacing}{\spaceskip=\fontdimen2\font plus
\BIBentryALTinterwordstretchfactor\fontdimen3\font minus
  \fontdimen4\font\relax}
\providecommand{\BIBforeignlanguage}[2]{{%
\expandafter\ifx\csname l@#1\endcsname\relax
\typeout{** WARNING: IEEEtran.bst: No hyphenation pattern has been}%
\typeout{** loaded for the language `#1'. Using the pattern for}%
\typeout{** the default language instead.}%
\else
\language=\csname l@#1\endcsname
\fi
#2}}
\providecommand{\BIBdecl}{\relax}
\BIBdecl

\bibitem{viterbi1967error}
A.~Viterbi, ``Error bounds for convolutional codes and an asymptotically
  optimum decoding algorithm,'' \emph{IEEE transactions on Information Theory},
  vol.~13, no.~2, pp. 260--269, 1967.

\bibitem{takamoto2017computing}
A.~Takamoto, M.~Yoshida, K.~Umemura, and Y.~Ichikawa, ``Computing information
  quantity as similarity measure for music classification task,'' in
  \emph{Proceedings of the 2017 International Conference on Advanced
  Informatics, Concepts, Theory, and Applications (ICAICTA)}, 2017.

\bibitem{lin2003symbolic}
J.~Lin, E.~Keogh, S.~Lonardi, and B.~Chiu, ``A symbolic representation of time
  series, with implications for streaming algorithms,'' in \emph{Proceedings of
  the 8th ACM SIGMOD workshop on Research issues in data mining and knowledge
  discovery}, 2003, pp. 2--11.

\bibitem{Chen2015}
Y.~Chen, E.~Keogh, B.~Hu, N.~Begum, A.~Bagnall, A.~Mueen, and G.~Batista, ``The
  ucr time series classification archive,'' July 2015,
  \url{www.cs.ucr.edu/~eamonn/time_series_data/}.

\bibitem{keogh2004towards}
E.~Keogh, S.~Lonardi, and C.~A. Ratanamahatana, ``Towards parameter-free data
  mining,'' in \emph{Proceedings of the tenth ACM SIGKDD international
  conference on Knowledge discovery and data mining}, 2004, pp. 206--215.

\bibitem{manber1993suffix}
U.~Manber and G.~Myers, ``Suffix arrays: a new method for on-line string
  searches,'' \emph{siam Journal on Computing}, vol.~22, no.~5, pp. 935--948,
  1993.

\bibitem{weiner1973linear}
P.~Weiner, ``Linear pattern matching algorithms,'' in \emph{Proceedings of the
  14th Annual Symposium on Switching and Automata Theory (swat 1973)}, 1973,
  pp. 1--11.

\bibitem{umemura2018non}
K.~Umemura, Y.~Kohara, N.~Yusuk, A.~Takamoto, and M.~Yoshida, ``Non-overlapping
  counting of string using suffix array,'' in \emph{Proceedings of the 2018 5th
  International Conference on Advanced Informatics: Concept Theory and
  Applications (ICAICTA)}, 2018, pp. 25--29.

\bibitem{brodalsolving}
G.~S. Brodal, R.~B. Lyngs{\o}, A.~{\O}stlin, and C.~N. Pedersen, ``Solving the
  string statistics problem in time {$O(n\log n)$},'' in \emph{Automata,
  Languages and Programming}, 2002, pp. 728--739.

\end{thebibliography}
\end{document}